\documentclass[aps,pre,groupedaddress,10pt,showkeys,nofootinbib,twocolumn]{revtex4-2}

\usepackage{graphics,graphicx}
\usepackage{amsmath,amssymb}
\usepackage{multirow}
\usepackage{color}
\usepackage{hyperref}

\usepackage[normalem]{ulem}  

\newcommand{\stat}{\mathrm{stat}}
\newcommand{\coal}{\mathrm{coal}}
\newcommand{\dr}{\mathrm{d}}

\begin{document}

\title{New nonet scalar mesons and glueballs:
the mass spectra and the production yields in relativistic heavy ion collisions}

\date{\today}

\author{Shigehiro Yasui}
\email[]{s-yasui@nishogakusha-u.ac.jp}
\email[]{yasuis@keio.jp}
\affiliation{Nishogakusha University, 6-16, Sanbancho, Chiyoda, Tokyo 102-8336, Japan}
\affiliation{International Institute for Sustainability with Knotted Chiral Meta Matter (SKCM$^2$), Hiroshima University, 1-3-2 Kagamiyama, Higashi-Hiroshima, Hiroshima 739-8511, Japan}
\affiliation{Research and Education Center for Natural Sciences, Keio University, Hiyoshi 4-1-1, Yokohama, Kanagawa 223-8521, Japan}
\author{Su Houng Lee}
\email[]{suhoung@yonsei.ac.kr}
\affiliation{Department of Physics and Institute of Physics and Applied Physics, Yonsei University, Seoul 03722, Korea}
\author{Pok Man Lo}
\email[]{pokman.lo@uwr.edu.pl}
\affiliation{Institute of Theoretical Physics, University of Wroclaw,
PL-50204 Wroc\l aw, Poland}
\author{Chihiro Sasaki}
\email[]{chihiro.sasaki@uwr.edu.pl}
\affiliation{Institute of Theoretical Physics, University of Wroclaw,
PL-50204 Wroc\l aw, Poland}
\affiliation{International Institute for Sustainability with Knotted Chiral Meta Matter (SKCM$^2$), Hiroshima University, 1-3-2 Kagamiyama, Higashi-Hiroshima, Hiroshima 739-8511, Japan}

\begin{abstract}
We propose a new nonet scheme for scalar mesons consisting of $f_{0}(980)$, $a_{0}(980)$, $K_{0}^{\ast}(1430)$, and $f_{0}(1770)$, regarding them as quark-antiquark $P$-wave states classified by $\mathrm{SU}(3)$ light flavor symmetry.
We investigate their production in relativistic heavy ion collisions, and estimate their yields by applying the statistical model and the quark coalescence model.
In contrast to these scalar mesons, we regard $f_{0}(1500)$ as a glueball that is not included in the proposed nonet.
We quantify the production yields of $f_{0}(1500)$ 
by accounting for various internal structures of this state,
and compare their yields with those of the new nonet scalar mesons.
Given the production yields of these hadrons, our results strongly suggests that $f_{0}(1500)$ is a glueball.
\end{abstract}

\maketitle


\section{Introduction}
\label{sec:introduction}

The scalar mesons 
with light flavors
have received considerable attention to date
since early days of the hadron spectroscopic study~\cite{ParticleDataGroup:2024cfk}.
Their studies are deeply concerned to unveil the confinement properties in the Quantum Chromodynamics (QCD), as they exhibit various exotic structures of hadrons, e.g., multiquarks, hadronic molecules, glueballs, and hybrids.

Since early days, the low-lying scalar mesons, such as $f_{0}(500)$ (isosinglet), $K_{0}^{\ast}(700)$ (isodoublet), $f_{0}(980)$ (isosinglet), and $a_{0}(980)$ (isotriplet), have been known as the candidates of exotic hadrons.
Indeed, 
their mass ordering,
$m_{f_{0}(500)} < m_{K_{0}^{\ast}(700)} < m_{f_{0}(980)} \simeq m_{a_{0}(980)}$, observed in experiments
remains elusive
in the simple quark model, i.e., quark-antiquark ($\bar{q}q$) bound states with relative $P$-wave~\cite{Jaffe:1976ig,Jaffe:1976ih}.
In the two-body quark-antiquark picture,
the lowest states should be $\bar{u}u+\bar{d}d$ and $\bar{u}u-\bar{d}d$, $\bar{d}u$ and $\bar{u}d$, the first excited states should be $\bar{s}u$ and $\bar{s}d$, and the second excited state should be $\bar{s}s$.
This naive expectation is due to an explicit breaking of the $\mathrm{SU}(3)$ light flavor symmetry stemming from the quark mass differences.
However, such a mass ordering is not consistent with past measurements.
Motivated to resolve this discrepancy, 
various exotic structures have been proposed:
compact tetraquarks ($\bar{q}^{2}q^{2}$)~\cite{Jaffe:1976ig,Jaffe:1976ih}, extended hadronic molecules ($\pi\pi$, $\pi K$, $K\bar{K}$), and others~\cite{Close:2002zu,Amsler:2004ps,Bugg:2004xu,Klempt:2007cp,Pelaez:2015qba,Pelaez:2021dak} (see also Sec.~62 in Ref.~\cite{ParticleDataGroup:2024cfk} for more recent progress).

In the literature,
some scalar mesons can be described as glueballs, i.e., the color-singlet objects 
(almost) purely composed
of gluons with possibly small admixture of quarks and antiquarks (see, e.g., Ref.~\cite{Crede:2008vw} for an experimental review).
Thus, studying glueballs is important for unveiling the mechanisms of color confinement and mass generation in pure Yang-Mills theory/QCD.
One of the famous glueball candidates is an isosinglet scalar meson $f_{0}(1500)$ with mass $1522\pm22$~MeV and decay width $108\pm33$~MeV in $J^{PC}=0^{++}$~\cite{ParticleDataGroup:2024cfk}.
The quantum number $J^{PC}$ is given by spin $J$, parity $P$ and charge-conjugation parity $C$.
Also, there exist several glueball candidates, such as $f_{0}(1710)$ and $f_{0}(2470)$ in $0^{++}$, $f_{J}(2220)$ in $2^{++}$ or $4^{++}$.
As a notable recent highlight,
$X(2370)$ was observed in high-energy experiments by BESSIII~\cite{BESIII:2023wfi,She:2024ewy,Liu:2024anq}.
Glueballs have been investigated in a variety of theoretical approaches, such as lattice QCD~\cite{Johnson:1997ap,Michael:1988jr,Bali:1993fb,Sexton:1995kd,Morningstar:1999rf,Johnson:2000qz,Liu:2001wqa,Meyer:2004jc,Chen:2005mg,Loan:2006gm,Gregory:2012hu,Brett:2019tzr,Athenodorou:2020ani,Chen:2021dvn,Sakai:2022zdc,Morningstar:2024vjk}, holographic QCD~\cite{Witten:1998zw,Sakai:2004cn}, and phenomenological models (e.g., flux-tube model~\cite{Isgur:1984bm,Kuti:1998rh}, bag model~\cite{Jaffe:1975fd}, constituent gluon model~\cite{Barnes:1981ac,Cornwall:1982zn}).
Glueballs can also be described as topological solitons, known as Hopfions, leading to mass spectra consistent with experiments and lattice QCD studies~\cite{Faddeev:1996zj,Faddeev:1998eq,Faddeev:2003aw,Kondo:2006sa,Amari:2025cre}.

In this paper, we consider the heavier scalar mesons, i.e., $f_{0}(980)$, $a_{0}(980)$, $K_{0}^{\ast}(1430)$ (isodoublet), and $f_{0}(1770)$ (isosinglet),
by regarding them as belonging to a novel light-flavor nonet.
This new nonet assignment is quite different from the conventional one, where
$f_{0}(500)$, $K_{0}^{\ast}(700)$, $f_{0}(980)$, and $a_{0}(980)$ are supposed to form a nonet.
In our new scheme, we naturally assign $f_{0}(1500)$ to a glueball.
In order to identify those classifications by experiments, we propose to study their production yields in relativistic heavy ion collisions (HICs)
at the Relativistic Heavy Ion Collider (RHIC) and the Large Hadron Collider (LHC).

It is noted that the ALICE experiment at the LHC has recently reported an intriguing data in $p$-Pb collisions, 
revealing the suppression of $f_{0}(980)$ productions against $K^{\ast0}(892)$ at transverse-momentum ($p_{T}$)
 up to about 4~GeV/$c$~\cite{ALICE:2023cxn}.
This result indicates that $f_{0}(980)$ contains only a small amount of $\bar{s}s$ fractions, and hence that $f_{0}(980)$ can be {\it a conventional meson with no hidden strange quarks}~\cite{ALICE:2023cxn} 
contrary to the compact tetraquark or molecular descriptions.
The original QCD sum rules
also find
that the masses of both $f_{0}(980)$ and $a_{0}(980)$ are  well reproduced when assuming the interpolating currents to be composed of the light quark-antiquark~\cite{Reinders:1984sr}, although alternative configurations such as tetraquark or mixed states are also viable~\cite{Chen:2006hy}.
Thus, 
this highlights the importance of reconsidering
the flavor structure of scalar mesons including $f_{0}(980)$, as it will be be discussed in detail in our new nonet scheme.

Furthermore, 
HICs provide a unique laboratory to study productions of exotic hadrons under extreme conditions~\cite{ExHIC:2010gcb,ExHIC:2011say,ExHIC:2017smd}.
In HICs, multiple
hadrons
 are produced abundantly by accompanying various flavors through  coalescence processes.
For example, one of the well-known exotic hadrons, $X(3872)$, which has been observed in $e^{-}e^{+}$ collisions~\cite{Belle:2003nnu,BaBar:2004oro} and also in $p\bar{p}$ and $pp$ collisions~\cite{CDF:2003cab,D0:2004zmu}, is now studied in HICs~\cite{CMS:2021znk}.
Because the inner structures of  $f_{0}(980)$, $a_{0}(980)$, $K_{0}^{\ast}(1430)$, and $f_{0}(1770)$ can be relevant to the coalescence processes, it is important to explore them in HICs.

The paper is organized as follows:
In Sec.~\ref{sec:new_nonet_scalar_mesons}, we propose the light-flavor nonet structure for $f_{0}(980)$, $a_{0}(980)$, $K_{0}^{\ast}(1430)$, and $f_{0}(1770)$, while we identify $f_{0}(1500)$ as a glueball.
In Sec.~\ref{sec:HIC_productions}, we estimate the production yields of $f_{0}(980)$, $a_{0}(980)$, $K_{0}^{\ast}(1430)$, $f_{0}(1770)$, and $f_{0}(1500)$ in HICs by employing the statistical model and the coalescence model.
We also perform an analysis in terms of the S-matrix formulation.
The final section is devoted to our conclusions and future perspectives.
In Appendix~\ref{sec:S-matrix_calculations}, we summarize the details of the S-matrix formulation.
In Appendix~\ref{sec:discussions}, we discuss the possible uncertainty caused by systematic errors in our estimates.

\section{New nonet scalar mesons and glueballs}
\label{sec:new_nonet_scalar_mesons}

While the quark mass dependence on flavors gives characteristic mass splittings among the multiplet number of hadrons, the overall pattern in mass spectrum reflects the underlying $\mathrm{SU}(3)$ light flavor symmetry.
According to the Particle Data Group~\cite{ParticleDataGroup:2022pth}, there are a variety of scalar mesons below 2~GeV, as shown in Fig.~\ref{fig:250206_spectrum_scalar}.
In the literature, it has been considered that $f_{0}(500)$, $K_{0}^{\ast}(700)$, $f_{0}(980)$, and $a_{0}(980)$ compose an octet of the $\mathrm{SU}(3)$ light flavor symmetry.
In the present study, we remove the some low-mass states, i.e., $f_{0}(500)$ and $K_{0}^{\ast}(700)$, from the list of  scalar multiplets, because they have large decay widths more than hundred MeV, and hence they are much unstable.
There are theoretical arguments~\cite{sigma1, sigma2, kappa} against the inclusion of the $f_{0}(500)$ ($\sigma$), and similarly the $K_{0}^{\ast}(700)$ ($\kappa$), as standard states in hadronization models.
Instead, we focus on $f_{0}(980)$, $a_{0}(980)$, $K_{0}^{\ast}(1430)$, and $f_{0}(1770)$, as indicated Fig.~\ref{fig:250206_spectrum_scalar},
and propose a new nonet structure.
Their nonet structure is shown in Fig.~\ref{fig:231109_nonet}~(a) also.
This is akin to the famous pattern of vector mesons, i.e., $\omega(782)$, $\rho(770)$, $K^{\ast}(892)$, and $\phi(1020)$, as shown in Fig.~\ref{fig:231109_nonet}~(b).

\begin{figure}[t]
\begin{center}
\includegraphics[scale=0.36]{figure_240224/250206_spectrum_scalar.pdf}
\caption{Mass spectra
of scalar mesons with nonstrangeness and strangeness
are shown
in the mass region below 2~GeV. The new nonet scalar mesons, $f_{0}(980)$, $a_{0}(980)$, $K_{0}^{\ast}(1430)$, and $f_{0}(1770)$, proposed in this study are indicated by arrows. The glueball candidate, $f_{0}(1500)$, is shown by dashed line. Cf.~Fig.~\ref{fig:231109_nonet}.}
\label{fig:250206_spectrum_scalar}
\end{center}
\end{figure}

The proposed nonet scalar mesons represent the quark-antiquark two-body systems in relative $P$-wave with light flavor structures, such as $\bar{u}u\pm\bar{d}d$ and $\bar{s}s$, as shown in Fig.~\ref{fig:231109_nonet}~(a).
In this nonet scheme, the quark-antiquark content of $f_{0}(980)$, $\bar{u}u+\bar{d}d$,  is quite different from the tetraquark picture, $\bar{u}\bar{s}us+\bar{d}\bar{s}ds$.
This can be supported by the recent study in ALICE collaboration suggesting that $f_{0}(980)$ includes small amount of $\bar{s}s$ fractions~\cite{ALICE:2023cxn}.
Thus, we regard $f_{0}(980)$, together with its isospin partner $a_{0}(980)$, being described by quark-antiquark picture: $f_{0}(980)$ by $\bar{u}u+\bar{d}d$, $a_{0}^{0}(980)$ by $\bar{u}u-\bar{d}d$, and $a_{0}^{\pm}(980)$ by $\bar{d}u$ and $\bar{u}d$, respectively.
As for $f_{0}(980)$ and $f_{0}(1770)$,
we suppose an ideal mixing stemming from the different masses for up and down quarks and a strange quark ($m_{u}, m_{d} \neq m_{s}$).
Thus, 
$f_{0}(1770)$ 
is described by $\bar{s}s$.
This mimics the the ideal mixing in the vector mesons, $\omega$ and $\phi$, as shown in Fig.~\ref{fig:231109_nonet}~(b).
The properties of
light hadrons discussed in the present study are summarized
in Table~\ref{tbl:summary_light_hadrons}.

\begin{figure}[t]
\begin{center}
\vspace{0em}
\includegraphics[scale=0.24]{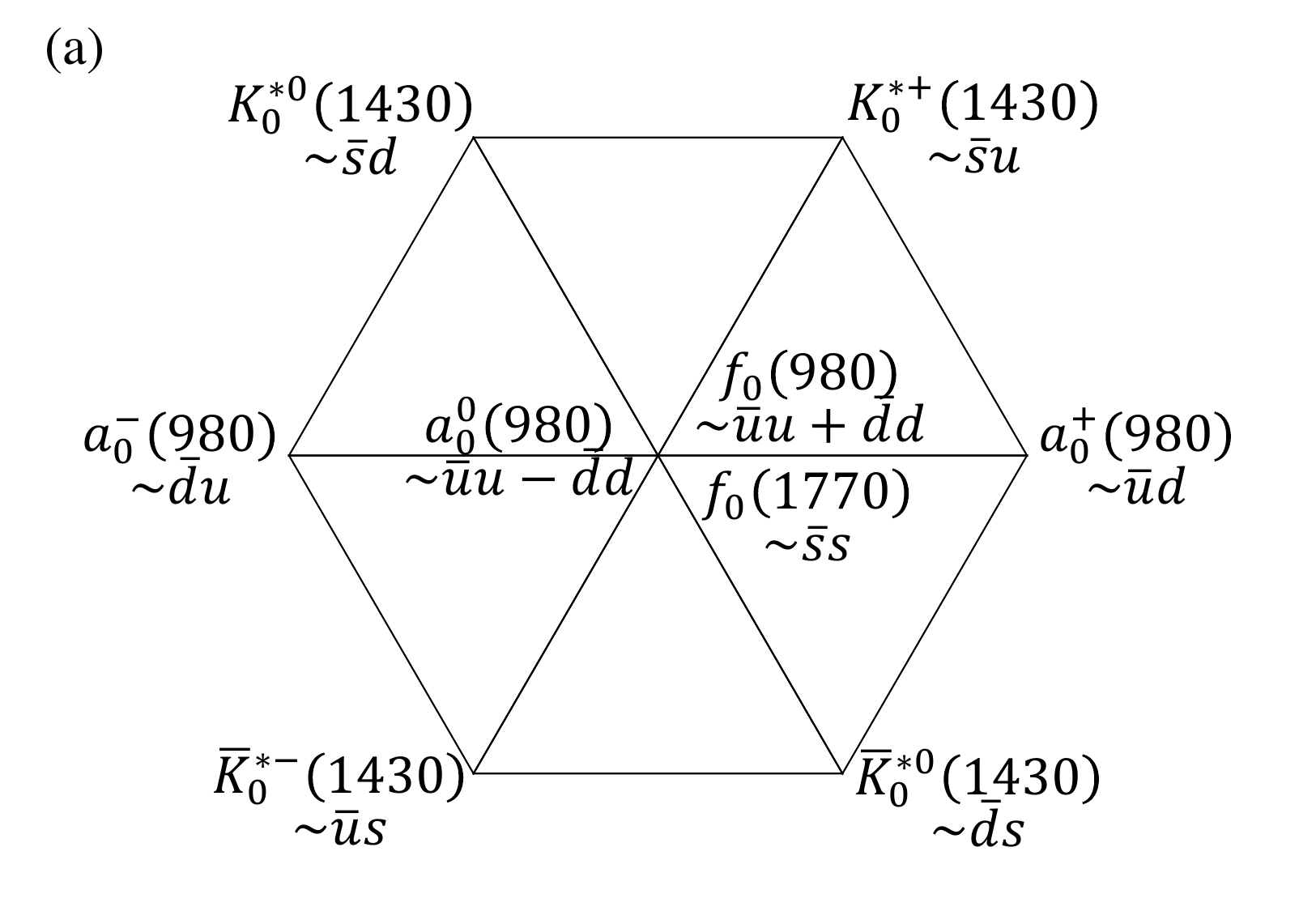}
\includegraphics[scale=0.24]{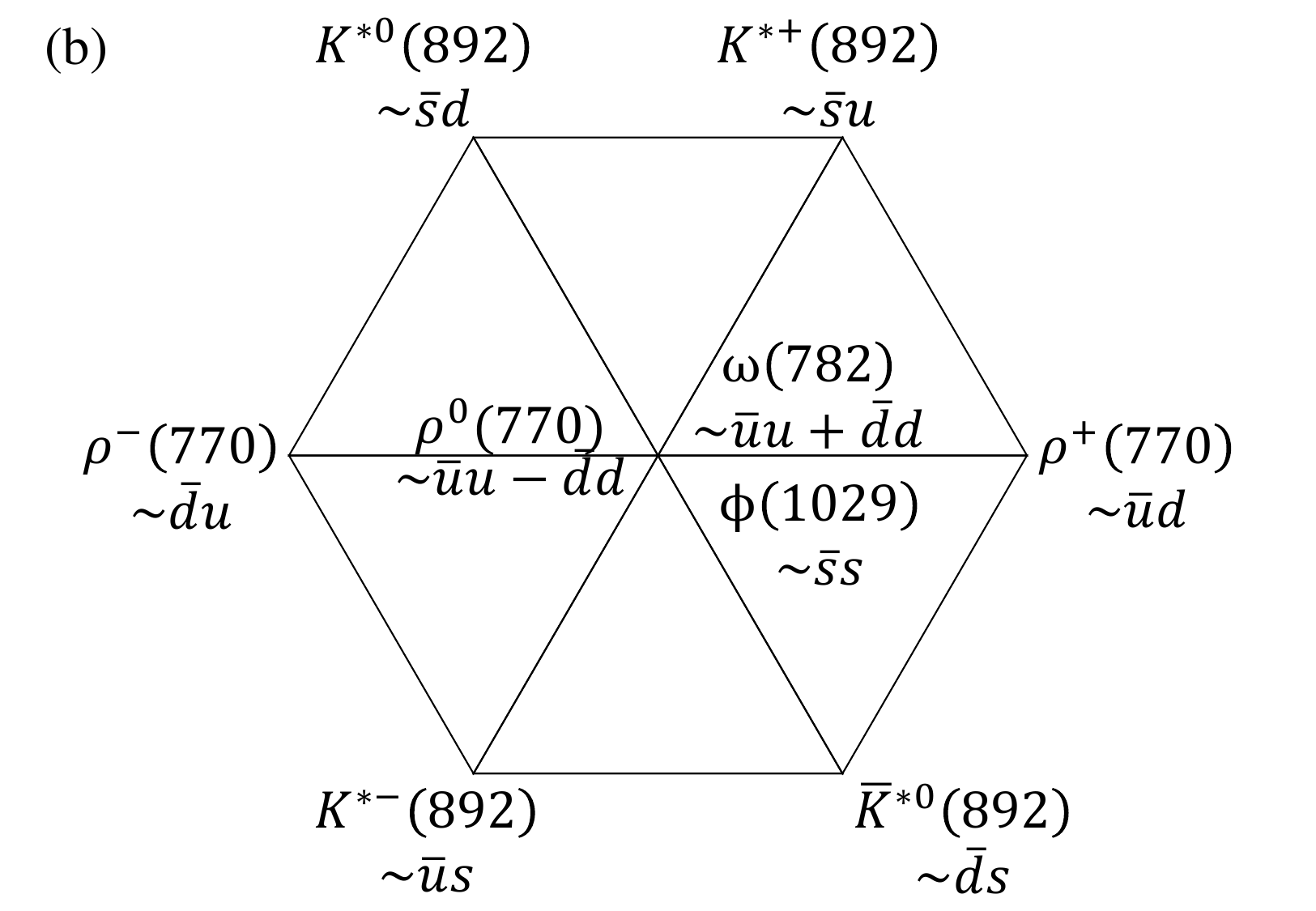}
\caption{The $\mathrm{SU}(3)$ flavor diagrams are shown for (a) new nonet scalar mesons proposed in this study (cf.~Fig.~\ref{fig:250206_spectrum_scalar}) and
(b) ordinary vector mesons.}\label{fig:231109_nonet}
\end{center}
\end{figure}

There are several remarks.
Firstly, because $f_{0}(1770)$ contains $\bar{s}s$ as a main component, we may consider that $f_{0}(1770)$ will decay predominantly to $K\bar{K}$ channel with decay width larger than those to the other light hadrons.
This should be expected because the analogue vector meson, $\phi(1020)$ accompanying with $\bar{s}s$ content, decays to $K\bar{K}$ with a large branching ratio (B.R. 50~\%).
Secondly, we exclude $f_{0}(1710)$ and $a_{0}(1710)$
from our consideration of the proposed nonet mesons.
This is simply because they cannot be naturally fit to the mass hierarchy in the new nonet scheme, when they are assigned to isospin partners like $f_{0}(980)$ and $a_{0}(980)$. 
It would be rather expected that $f_{0}(1710)$ and $a_{0}(1710)$, together with $K_{0}^{\ast}(1950)$, should belong to another nonet with heavier masses.
If this is the case, the candidate with $\bar{s}s$ would be assigned to $f_{0}(2100)$ or $f_{0}(2200)$.
More studies of such a possible second class of nonet scalar mesons are postponed.
Thirdly, $a_{0}(1450)$ is outside of our scope because it is regarded to be much unstable due to its large decay width.

Although $f_{0}(1380)$ and $f_{0}(1500)$ lie inside the mass range of the new nonet scalar mesons as quark-antiquark systems,
we consider them to be excluded from the proposed nonet pattern.
We rather regard them as exotic states: $f_{0}(1380)$ as 
a state generated kinematically by rescatterings, and $f_{0}(1500)$ as a glueball, i.e., a bound state of two gluons ($gg$).
We exclude $f_{0}(1380)$ from the present study,
because the generation mechanism of $f_{0}(1380)$ cannot be described within the coalescence model.
We concentrate on $f_{0}(1500)$ only.
We may know that the other possible candidates of glueballs with heavier masses, such as $f_{0}(2020)$, $f_{0}(2100)$, $f_{0}(2470)$, and so on, are found in the list of Particle Data Group~\cite{ParticleDataGroup:2022pth}.
They indeed can provide interesting inner structures.
For example, $f_{0}(2470)$ may be described as a glueball by the Hopfion approach.
Nevertheless, we postpone the analysis for heavier possible glueballs to future studies.
Instead, we focus on $f_{0}(1500)$ and investigate its production, not only as a glueball, but also as a quark-antiquark state or a multiquark state.
Such comparison for different structures will clarify the particular properties of the glueball productions.

\begin{table}[t]
\caption{Summary of normal and exotic hadrons with light flavors for the new nonet, the glueball candidates, and the reference particles. The mass ($m$), decay width ($\Gamma$), isospin ($I$), spin and parity
($J^P$) are shown~\cite{ParticleDataGroup:2022pth}. The possible quark and gluon structures ($\bar{q}q$, $qqq$, $\bar{q}^{2}q^{2}$, and $gg$ with internal angular momenta $L=S,P$) are presented in the last column (confs.).
The notations $q=u,d,s$ for light quarks and $n=u,d$ for non-strange light quarks are used.}\label{tbl:summary_light_hadrons}
\begin{center}
\begin{tabular}{l|rrrrr}
\hline \hline
& particle & $m$ [MeV] & $\Gamma$ [MeV] & $I(J^{P})$ & confs.($L$) \\
\hline \hline
 nonet & $f_{0}(980)$ & $990\pm20$ & 10-100 & $0(0^{+})$ & $\bar{n}n$($P$) \\
 & $a_{0}(980)$ & $980\pm20$ & 50-100 & $1(0^{+})$ & $\bar{n}n$($P$) \\ 
 & $K_{0}^{\ast}(1430)$ & $1425\pm50$ & $270\pm80$ & $\frac{1}{2}(0^{+})$ & $\bar{s}n$($P$) \\
 & $f_{0}(1770)$ & $1733_{-7}^{+8}$ & $150_{-10}^{+12}$ & $0(0^{+})$ & $\bar{s}s$($P$) \\
\hline
 glueball & $f_{0}(1500)$ & $1522\pm25$ & $108\pm33$ & $0(0^{+})$ & $\bar{n}n$($P$) \\
 & & & & & $\bar{n}^{2}n^{2}$($S$) \\
 & & & & & $gg$ ($S$) \\
\hline
 normal & $\rho(770)$ &775 & 147 & $1(1^{-})$ & $\bar{n}n$($S$) \\
 & $\phi(1020)$ & 1019 & 4.3 & $0(1^{-})$ & $\bar{s}s$($S$) \\
 & $\Omega(1672)$ & 1672& --- & $0(\frac{3}{2}^{+})$ & $sss$($S$) \\
\hline \hline
\end{tabular}
\end{center}
\end{table}%

\section{Production of scalar mesons and glueballs in heavy-ion collisions}
\label{sec:HIC_productions}


In this section, we study the production of the proposed nonet scalar mesons ($\bar{q}q$) and gluons ($gg$) in HICs.
We employ the statistical model as a standard approach.
In addition, we adopt the particle coalescence model,
to investigate the dependence of the yields on their internal structures.

\subsection{Statistical model}
\label{sec:stat}

Firstly, we consider the statistical model~\cite{Andronic:2005yp}, which provides the hadron yields according to the thermal distribution function at finite temperature.
In the statistical model, the essential inputs are only the hadron mass and the degree of degeneracy.
In spite of its simplicity, the statistical model has been successful in explaining the experimental yields of a variety of hadrons and nuclei, including hadron resonances, in HICs.
Thus, applying the statistical model to the new nonet scalar mesons and glueballs is a good starting point.

In the statistical model~\cite{Andronic:2005yp}, 
the yield of hadron $h$ is given by
\begin{align}
   N_{h}^{\stat}
= V_{H} \frac{g_{h}}{2\pi^{2}}
   \int_{0}^{\infty} \frac{p^{2} \dr p}{e^{(E_{h}-\mu_{B}B-\mu_{S}S)/T_{H}}\pm1},
\end{align}
for fermions and bosons ($\pm1$),
where $E_{h}=\sqrt{p^{2}+m_{h}^{2}}$ is the energy of hadron $h$ with mass $m_{h}$ and momentum $p$.
Here $g_{h}$ is the degree of degeneracy of hadron $h$.
$B$ ($S$) is the baryon (strangeness) numbers of hadron $h$, and $\mu_{B}$ ($\mu_{S}$) is the corresponding chemical potential.
$T_{H}$ and $V_{H}$ are hadronization temperature and volume, respectively, of the source region for producing hadron $h$, where the statistical production occurs.

As for the values of statistical parameters, we follow the previous analyses applied for several exotic hadrons in Refs.~\cite{Andronic:2012dm,Stachel:2013zma},
and take $T_{H}=162$ (156)~MeV and $V_{H}=2100$ (5380)~fm$^{3}$ for RHIC (LHC).
We consider only the collision energy 200~GeV in RHIC,
and we suppose two different collision energies, 2.76~TeV and 5.02~TeV, in LHC. 
It is noted in our model setting that $T_{H}$ and $V_{H}$ are interrelated by the temperature dependence of the volume of the hot matter expanding isentropically, as discussed in Ref.~\cite{ExHIC:2017smd}.
The values of all the parameters are summarized in Table~\ref{tbl:parameters}.
These parameters can explain the productions of normal hadrons, i.e., $\rho(770)$ and $\phi(1020)$ mesons as quark-antiquark states and $\Omega(1670)$ baryons as three-quark states (cf.~Table~\ref{tbl:summary_light_hadrons}).
They play the role of reference particles, because their internal structures are simpler.
The same values for parameters were applied to the exotic hadron productions from HICs~\cite{ExHIC:2010gcb,ExHIC:2011say,ExHIC:2017smd}, in which the freeze-out conditions analyzed in Ref.~\cite{Cho:2015exb} were considered.

\subsection{Coalescence model}
\label{sec:coalescence}

Secondly, 
we consider the coalescence model by taking into account the internal structures of
the constituent particles in the final state hadrons.
To achieve this,
we need to introduce the wave functions for the bound systems of quark-antiquark ($\bar{qq}$) and gluon-gluon ($gg$),
which are convoluted with the thermal distribution functions of the constituent particles (quarks, antiquarks, and gluons) at finite temperature.
To construct the wave functions, we follow the conventional nonrelativistic Schr\"odinger equation with a color confinement force expressed by the harmonic oscillator potential~\cite{ExHIC:2010gcb,ExHIC:2011say,ExHIC:2017smd}.
We regard the glueball as being made of two constituent gluons ($gg$) with a dynamical mass parametrized by $m_{g}$.

In the coalescence model, the final-state hadron $h$ is described by the composite states made of $n$ constituent particles.
The constituents are labeled by $i=1,2,\dots,n$.
The coalescence process is formalized by the overlapping of the Wigner functions of the composite states and the thermal distribution functions of the constituent particles.
The Wigner functions, representing the internal structures of the composite states, are expressed by the wave functions in the harmonic oscillator potential with an angular momentum frequency $\omega_{h}$ for the final-state hadron $h$.

With the above setting, the production yield of hadron $h$ is expressed by
\begin{align}
   N_{h}^{\coal}
\simeq
   g'_{h}
   \prod_{i=1}^{n-1} \frac{(4\pi\sigma_{i}^{2})^{3/2}}{V_{C}(1+2\mu_{i}T_{C}\sigma_{i}^{2})}
   \biggl( \frac{4\mu_{i}T_{C}\sigma_{i}^{2}}{3(1+2\mu_{i}T_{C}\sigma_{i}^{2})} \biggr)^{l_{i}},
\label{eq:yield_coalescence}
\end{align}
with $g'_{h}=g_{h} \prod_{j=1}^{n} ({N_j}/{g_j})$, where $g_{h}$ is the number of degeneracy for hadron $h$.
For the  $i$th constituent, $g_{i}$ is the degeneracy, $l_{i}=0,1,\dots$ indicate $S$, $P$, $\dots$ waves in the harmonic oscillator potential, and $N_{i}$ is the number of the constituent.
$T_{C}$ ($V_{C}$) is a temperature (volume) for the thermal distribution of the constituents.
The parameter $\sigma_{i}=1/\sqrt{\mu_{i}\omega_{h}}$ indicates the size of the wave function, where $\mu_{i}$ is the reduced mass defined by $\mu_{i}^{-1}=m_{i+1}^{-1}+(\sum_{j=1}^{i}m_{j})^{-1}$ with the $i$th constituent mass $m_{i}$.
The values of masses of up, down, and strange quarks are given by $m_{u}=m_{d}=350$~MeV and $m_{s}=500$~MeV, respectively~\cite{ExHIC:2017smd}.
In Eq.~\eqref{eq:yield_coalescence}, we have adopted the non-relativistic approximation, have neglected the transverse flow of produced matter, and have considered the central unit rapidity only (see, e.g., Ref.~\cite{ExHIC:2017smd}).

The critical temperature $T_{C}$ and volume $V_{C}$
are in principle determined by the entropy conservation in the interval from the critical time to the hadronization time~\cite{ExHIC:2017smd}.
By following Ref.~\cite{ExHIC:2017smd}, 
we adopt two different scenarios for determining $T_{C}$ and $V_{C}$, 
in which
 different settings for the phase transition are considered.
The first scenario (Sc.1) is to take
$T_{C}=T_{H}$ and $V_{C}=V_{H}$.
In this scenario, it is assumed that hadrons are continuously produced during the interval time of the crossover phase transition from QGP to hadron gas.
With this assumption in Sc.1, we consider that the number of hadrons estimated by the statistical model should be equal to that estimated by the quark coalescence model.
In the second scenario (Sc.2), we alternatively determine $T_{C}$ so that the hadron yields calculated by the quark coalescence model at $T_{C}$ are equal to those calculated by the hadron statistical model at $T_{H}$.
In both scenarios (Sc.1 and Sc.2), the critical volume $V_{C}$ is uniquely determined by the temperature dependence of the volume of the matter during its isentropic expansion (see, e.g., Ref.~\cite{ExHIC:2017smd}).

\begin{table}[tbp] 
\caption{The parameters in the statistical and coalescence models in Sc.1 and Sc.2 at
RHIC (200~GeV), LHC (2.76~TeV, 5.02~TeV),
and those given in Ref.~\cite{ExHIC:2017smd}.
Non-strange (up and down) and strange quark masses are taken to be 
$m_{u}=m_{d}=350$~MeV and $m_{s}=500$~MeV, respectively.
The parameter sets used in Refs.~\cite{ExHIC:2010gcb,ExHIC:2011say} are also shown for reference.\footnote{Notice that in Refs.~\cite{ExHIC:2010gcb,ExHIC:2011say} non-strange (up and down) light quark masses were taken to be $m_n=300$~MeV which is different from $350$~MeV in Ref.~\cite{ExHIC:2017smd} and the present study.}
}\label{tbl:parameters}
\begin{tabular}{l|cc|cccc|cc}
\hline \hline
	& \multicolumn{2}{c|}{RHIC} & \multicolumn{2}{c}{LHC(2.76)} & \multicolumn{2}{c|}{LHC(5.02)}
	& RHIC & LHC \\
	& Sc.1 & Sc.2
	& Sc.1 & Sc.2 
	& Sc.1 & Sc.2
	& \multicolumn{2}{c}{Refs.~\cite{ExHIC:2010gcb,ExHIC:2011say}}\\
\hline \hline
$T_{H}$ [MeV]    &\multicolumn{2}{c|}{162} &\multicolumn{4}{c|}{156} & \multicolumn{2}{c}{175} \\
$V_{H}$ [fm$^3$]  &\multicolumn{2}{c|}{2100}&\multicolumn{4}{c|}{5380} & 1908 & 5152 \\ 
$\mu_{B}$ [MeV] &\multicolumn{2}{c|}{24}  &\multicolumn{4}{c|}{0} & 20 & 0 \\ 
$\mu_{s}$ [MeV] &\multicolumn{2}{c|}{10}  &\multicolumn{4}{c|}{0} & 10 & 0 \\ 
\hline
$T_{C}$ [MeV]		& 162  & 166 & 156  & 166 & 156  & 166 & \multicolumn{2}{c}{175} \\
$V_{C}$ [fm$^3$]		& 2100 & 1791 & 5380 & 3533 & 5380 & 3533 & 1000 & 2700 \\
$\omega_{n}$ [MeV]		& 590 & 608 & 564 & 609 & 564 & 609 & \multicolumn{2}{c}{550} \\
$\omega_{s}$ [MeV]	& 431 & 462 & 426 & 502 & 426 & 502 & \multicolumn{2}{c}{519} \\
$N_{u}=N_{d}$		& 318 & 300 & 695 & 593 & 695 & 593 & 245 & 662\\
$N_{s}=N_{\bar{s}}$	& 183 & 176 & 386 & 347 & 386 & 347 & 150 & 405\\
\hline
\end{tabular}
\end{table}

As for the internal structure of a hadron $h$, the oscillator frequency ($\omega_{h}$) in the Wigner function is responsible to the extension of the hadron wave function in matter.
Its value should be chosen appropriately so that the reference particles, e.g., $\rho(770)$ and $\phi(1020)$ mesons and $\Omega(1670)$ baryons, satisfy the matching conditions: the yields of normal hadrons from the coalescence model is the same as those from the statistical model~\cite{ExHIC:2010gcb,ExHIC:2011say}.
This condition stems from the observation that the statistical model is successful in calculating the production yields obtained in the experiments, and hence it should be regarded as a prototype model of the hadron productions in HICs.

We consider,
according to Refs.~\cite{ExHIC:2010gcb,ExHIC:2011say},
 that the value of $\omega_{h}$ depends
on the flavor content of hadron $h$.
For example, 
the values of $\omega_{h}$ for non-strange hadrons are determined by the matching condition for $\rho(770)$ mesons.
This value 
will be denoted by $\omega_{n}$
in the followings.
Similarly, the values of $\omega_{h}$ for strange hadron, which will be denoted by $\omega_{s}$, are determined by the matching condition for $\Lambda(1105)$ baryons, provided that the additional yields contributed from higher resonance decays need to be included as the feed-down effect from the excited baryons.
The same value of $\omega_{s}$ is adopted to $\phi(1020)$ and $\Omega(1672)$, where the matching conditions are satisfied successfully.
In any case, it should be emphasized that the $\omega_{h}$ takes different values ($\omega_{n}$ or $\omega_{s}$) depending on non-strangeness ($n$) or strangeness ($s$) of flavor content in the hadron.

\begin{figure}[tbp]
\begin{center}
\vspace{0em}
\includegraphics[scale=0.33]{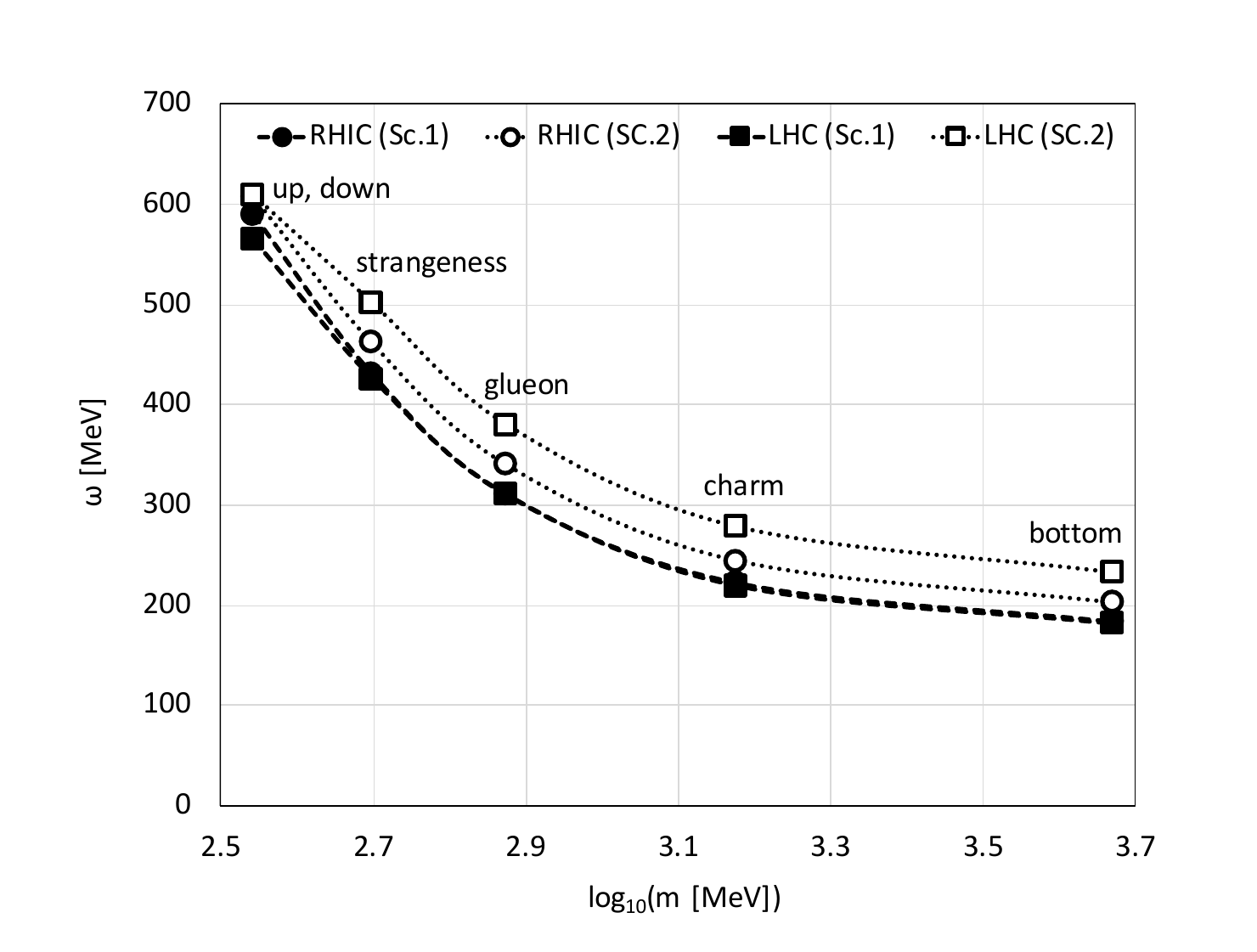}
\caption{The harmonic oscillator frequencies are plotted for the quark masses for up and down quarks ($u$ and $d$), strange quark ($s$), charm quark ($c$), and bottom quark ($b$).
They are denoted by $\omega_{n}$, $\omega_{s}$, $\omega_{c}$, and $\omega_{b}$ in the text.
The results shown for in the RHIC and LHC parameters, where the LHC results are the same for 2.76~TeV and  5.02~TeV.
Interpolating the points, the frequency $\omega_{g}$ for the constituent gluons ($g$) is obtained for the gluon mass $m_{g}=750$~MeV.
See Table~\ref{tbl:omega_g}.}
\label{fig:240223_omega_g}
\end{center}
\end{figure}

The values of $\omega_{n}$ and $\omega_{s}$ are summarized in Table~\ref{tbl:parameters}.
Including $\omega_{c}$ for charm hadrons and $\omega_{b}$ for bottom hadrons obtained in the previous studies~\cite{ExHIC:2017smd},
we plot the values of $\omega_{n}$, $\omega_{s}$, $\omega_{c}$, and $\omega_{b}$ for each quark mass in Fig.~\ref{fig:240223_omega_g}.
They are shown for Sc.1 and Sc.2 in the RHIC and LHC parameters.
We plot also the flavor dependence of $\omega_{h}$ on the quark mass in Fig.~\ref{fig:240223_omega_g}.
In the figure, the curves are fitted smoothly for the values of $\omega_{n}$, $\omega_{s}$, $\omega_{c}$, and $\omega_{b}$ for each scenario.
In this way, we obtain the dependence of $\omega_{h}$ on the mass of constituents.
We apply the obtained curves to the calculations in the coalescence model for the new nonet scalar mesons, i.e., $f_{0}(980)$, $a_{0}(980)$, $K_{0}^{\ast}(1430)$ and $f_{0}(1770)$, and the glueball candidate, i.e., $f_{0}(1500)$, by paying the following attentions.

\begin{table}[tbp]
\caption{The harmonic oscillator frequencies ($\omega_{g}$) and the numbers of gluons ($N_{g}$)
for the constituent gluons in Sc.1 and Sc.2 at RHIC (200~GeV), LHC (2.76~TeV, 5.02~TeV).
The gluon mass is set to be $m_{g}={750}$~MeV.}\label{tbl:omega_g}
\begin{tabular}{c|cc|cccc}
\hline \hline
	& \multicolumn{2}{c|}{RHIC} & \multicolumn{2}{c}{LHC(2.76)} & \multicolumn{2}{c}{LHC(5.02)} \\
	& Sc.1 & Sc.2
	& Sc.1 & Sc.2 
	& Sc.1 & Sc.2 \\
\hline \hline
 $\omega_{g}$ [MeV] & {310} & {340} & {310} & {380} & {310} & {380} \\
 $N_{g}$ &{248} & {248} & {497} & {489} & {497} & {489} \\
\hline
\end{tabular}
\end{table}

Firstly, we introduce the ansatz that the values of $\omega_{h}$ are dependent on the number of strange quarks in the scalar mesons.
Namely, instead of $\omega_{s}$, we introduce the enhanced angular momentum frequencies by some factors: $\omega_{s}' \approx 2\omega_{s}$ for $K_{0}^{\ast}(1430)$ and $\omega_{s}'' \approx 3\omega_{s}$ for $f_{0}(1770)$.
This ansatz is motivated by the following phenomenological consideration. In the coalescence model, $\omega_{s}$
is chosen such that the production yield of the ground-state hadron matches that obtained in the statistical model.
For excited states with non-zero relative angular momentum, however, the use of a Wigner function based on a Gaussian wave function does not realistically reflect the effects of the true interaction potential in the hadronic wave function.
In fact, Eq.~\eqref{eq:yield_coalescence}
indicates that the temperature-dependent suppression factor for non-zero $l_i$, is to leading order in
$\omega_{h}/T_{C}$,
given as
$1-\omega_{h} l_{i}/(2 T_{C})$.
This corresponds to the thermal suppression expected for a simple harmonic oscillator potential and is therefore not realistic for hadronic systems, where the excited states will see a more complicated long distance part of the potential.  
Consequently, compared to $K^{\ast}(892)$,  the $l=1$ excited state would acquire a temperature-dependent suppression factor corresponding to a hadron with an effective mass $892+\omega_{h}/2$.
Taking an approximate value of $\omega_{s} \approx 500$~MeV from Fig.~\ref{fig:240223_omega_g},
this would imply a statistical suppression similar to that of a hadron with a mass of about 1142 MeV, and hence unrealistic. This issue can be remedied by introducing an effective frequency 
$\omega_{s}' \approx 2 \omega_{s}$,
which produces a stronger thermal suppression consistent with that expected for a hadron with a mass close to 
that of $K_{0}^{\ast}(1430)$. An even larger multiplication factor,
$\omega_{s}'' \approx 3\omega_{s}$,
would be required when comparing the mass difference between  $\phi(1020) $ and $f_{0}(1770)$.
More discussions will be given in Appendix~\ref{sec:ws_dependence}.

Secondly, as for the flavor dependence of $\omega_{h}$ to a glueball candidate $f_{0}(1500)$, we suppose that the constituent-mass-dependence of $\omega_{h}$ in Fig.~\ref{fig:240223_omega_g} can be applied for constituent gluons.
Thus, from the extrapolated curved in Fig.~\ref{fig:240223_omega_g}, we estimate the value of $\omega_{h}$ for gluons, which will be denoted by $\omega_{g}$, by setting the mass of a constituent gluon, $m_{g}=750$~MeV.
This gluon mass is estimated by a half value of the $f_{0}(1500)$ mass.
The obtained values of $\omega_{g}$ are summarized for Sc.1 and Sc.2 in the RHIC and LHC parameters in Table~\ref{tbl:omega_g}.
In the table, the numbers of gluons obtained by the statistical model are shown also.

\begin{figure}[t]
\begin{center}
\begin{tabular}{cc}
\includegraphics[scale=0.19]{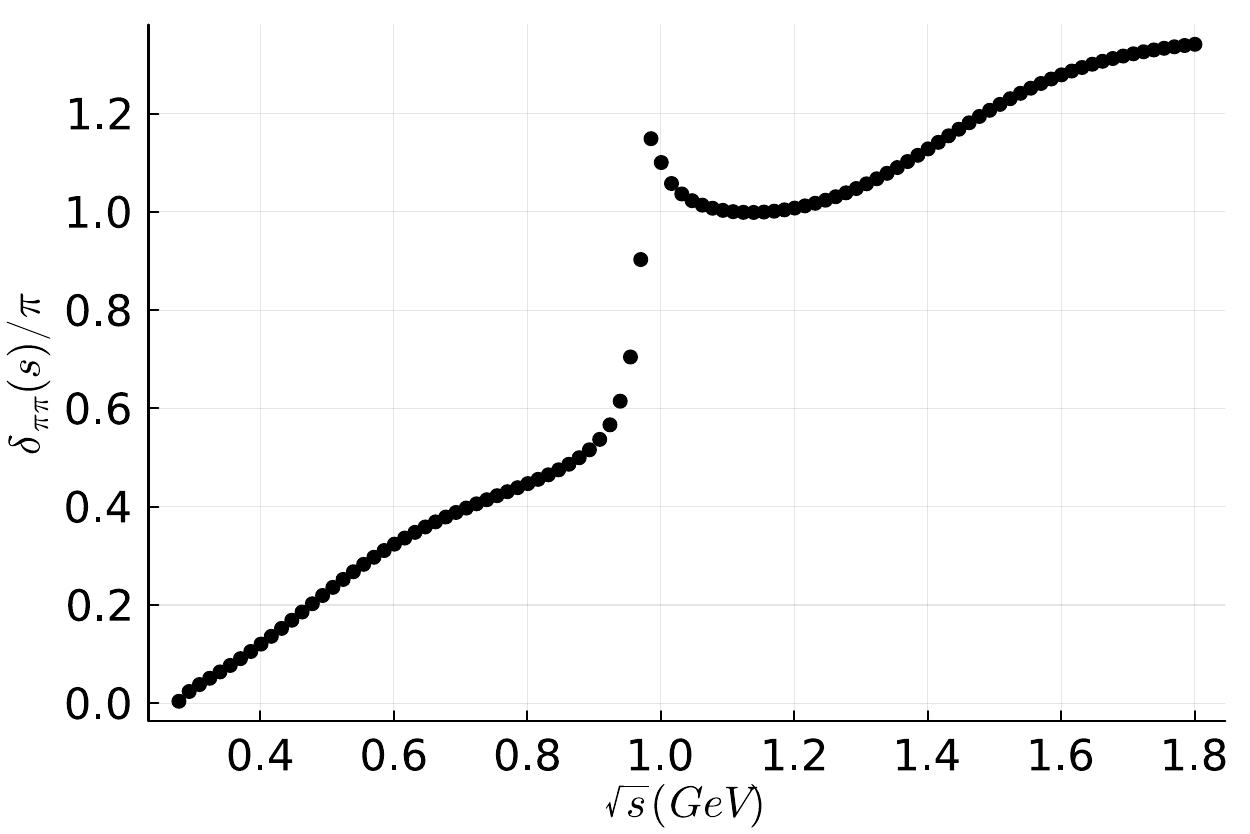} &
\includegraphics[scale=0.19]{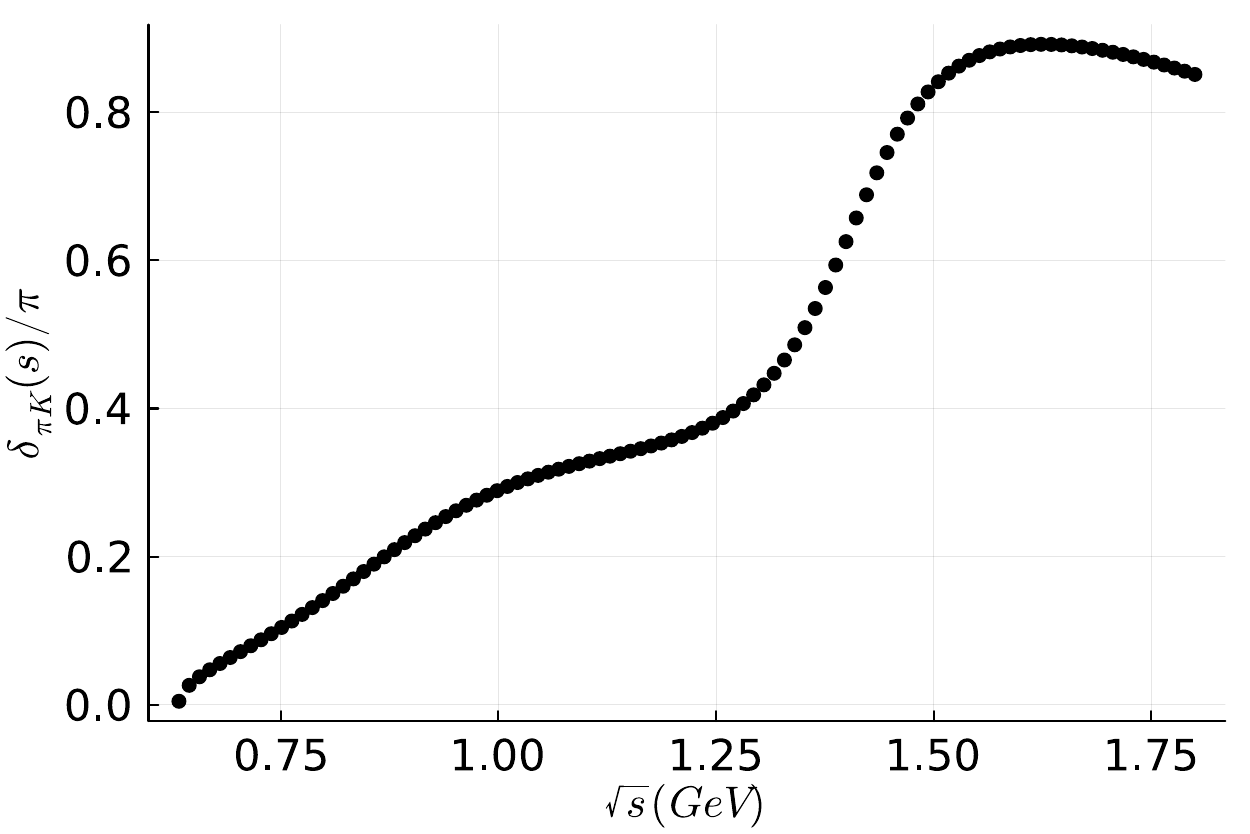} \\
{\small (a) $\pi\pi-K\bar{K}$} & \small{(b) $\pi K$}
\end{tabular}
\end{center}
\caption{The phase shifts~\cite{ccm,kappa} are shown to indicate (a) $f_{0}(500)$, $f_{0}(980)$, and  $f_{0}(1500)$ in $\pi \pi - K\bar{K}$ scatterings and (b) $K_{0}^{\ast}(700)$ and $K_{0}^{\ast}(1430)$ in $\pi K$ scatterings in the calculations in the S-matrix formulation.}
\label{fig:ps01_ps02}
\end{figure}

We also estimate the corresponding yields within the S-matrix formulation of statistical mechanics. In this approach, the thermal contribution of an interacting hadronic channel is determined directly from the scattering phase shift, whose energy derivative acts as an effective density of states and thus incorporates the full interaction dynamics. For this comparison, we employ the phase shifts from the $(\pi\pi, K\bar{K})$ coupled-channel analysis of Ref.~\cite{ccm} and $\pi K$ scattering from Ref.~\cite{kappa}. 
The S-matrix approach is in contrast to the statistical model and the coalescence model, because the latter two attribute thermal yields to individual states. Since the phase shift is defined on the physical energy axis, it contains both resonant and non-resonant contributions, which are not separable without additional modeling assumptions.
The phase shifts of $\pi \pi-K\bar{K}$ and $\pi K$ scatterings are presented in Fig.~\ref{fig:ps01_ps02}.

\begin{figure}[t]
\begin{center}
\vspace{0em}
\includegraphics[scale=0.28]{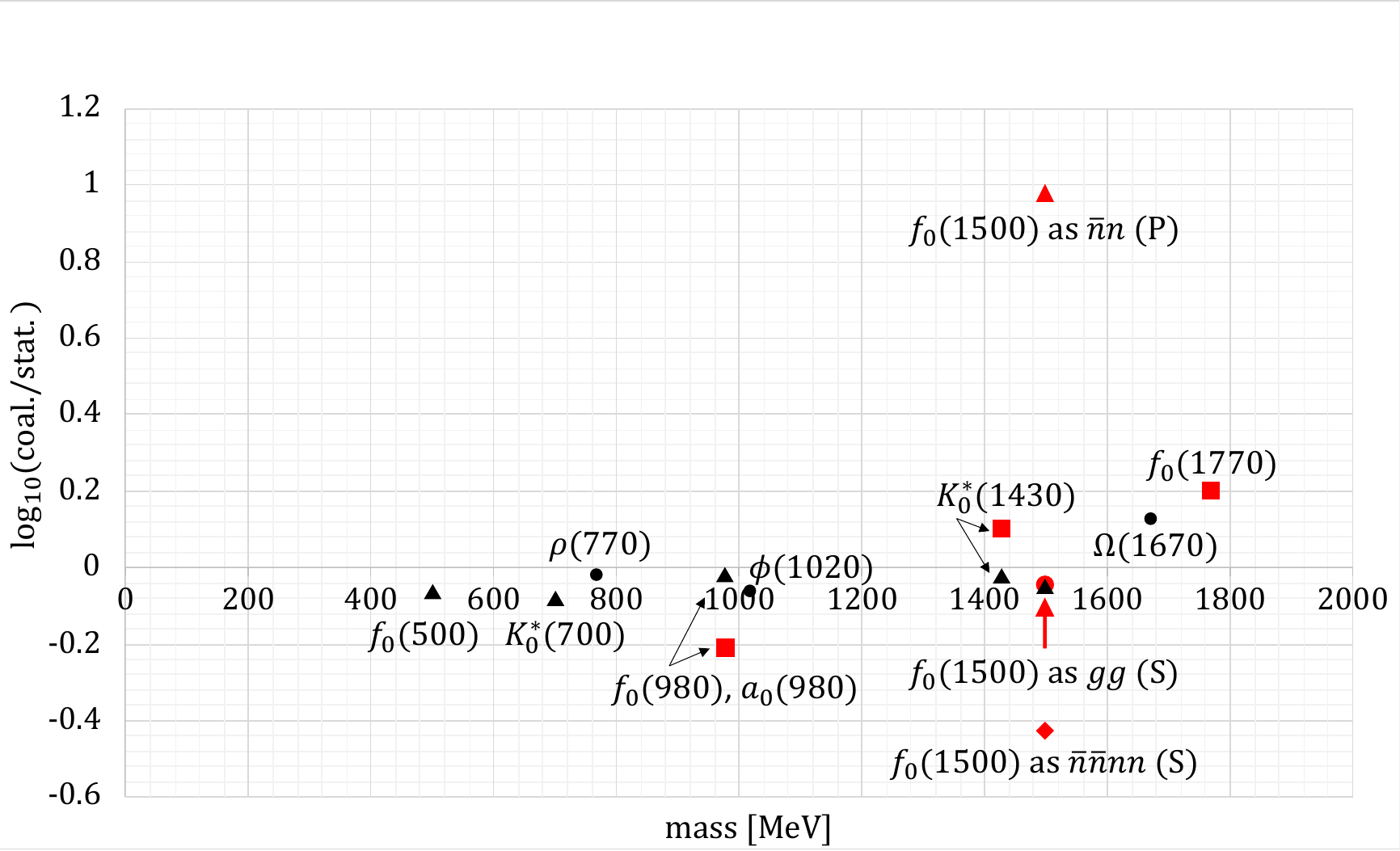}
\caption{The ratios of the hadron yields from the statistical model and the coalescence model (denoted by coal./stat.) are shown for the parameter set RHIC (Sc.1).
The normal hadrons are indicated by the black dots~($\bullet$).
The new nonet mesons, i.e., $f_{0}(980)$, $a_{0}(980)$, $K_{0}^{\ast}(1430)$, and $f_{0}(1770)$, are indicated by the red squares~({\color{red}$\blacksquare$}).
As for $f_{0}(1500)$, the $\bar{n}n$ ($P$-wave), $\bar{n}^{2}n^{2}$ ($S$-wave), and $gg$ ($S$-wave) configurations are indicated by the red triangle~({\color{red}$\blacktriangle$}), dot~({\color{red}$\bullet$}), and diamond~({\color{red}$\blacklozenge$}), respectively ($n=u,d$).
The ratios from the S-matrix formulation are shown by black-triangles~({\color{black}$\blacktriangle$}).
See Table~\ref{tbl:numerical_results} also.}\label{fig:250108_RHIC_Sc1_ratio}
\end{center}
\end{figure}

To enable a direct comparison, we evaluate the S-matrix contribution over the invariant-mass interval associated with the state of interest, as summarized in Table~\ref{tbl:numerical_results}. The corresponding systematic uncertainty is estimated by varying the chosen energy window.
The details of the S-matrix calculations are summarized in Appendix~\ref{sec:S-matrix_calculations}.
The estimated yield ratios relative to statistical models are 0.82 for $f_{0}(980)$, 0.57 for $f_{0}(1500)$, and 0.79 for $K_{0}^{\ast}(1430)$. 
For reference, we also obtain 0.48 for $f_{0}(500)$ and 0.38 for $K_{0}^{\ast}(700)$ as low-lying scalar mesons.
Such broad states are calculable in the S-matrix approach, because full-phase shifts in $\pi\pi$ and/or $\pi K$ scatterings are considered.
As a result, we will find that those numbers are comparable with the ratios of the coalescence model to the statistical model (coal./stat.), as it will be shown in Fig.~\ref{fig:250108_RHIC_Sc1_ratio} (cf.~Sec.~\ref{sec:ratios_hadron_yields}).
Thus, it will turn out that the yield estimates of resonant states in the statistical model and the coalescence model are not so different from those in the S-matrix approach.

\subsection{Numerical results}
\label{sec:results}

\begin{table}[tbp]
\caption{The yields of normal and exotic hadrons in the statistical model (stat.) and the coalescence model (coal.) for Sc.1 and Sc.2 at
RHIC (200~GeV), LHC (2.76~TeV, 5.02~TeV), respectively.
Notice that $\omega_{s}$ in normal $S$-wave hadrons is changed to $\omega_{s}' \approx 2\omega_{s}$ for $K_{0}^{\ast}(1430)$ and to $\omega_{s}'' \approx 3\omega_{s}$ for $f_{0}(1770)$ in the nonet scalar mesons.
The results by the S-matrix formulation (smat.) are shown only for RHIC, where the uncertainties are indicated in the parentheses.
See Fig.~\ref{fig:250108_RHIC_Sc1_ratio} also.}\label{tbl:numerical_results}
\begin{tabular}{r|r|r|cc|cccc}
\hline \hline
              &
             particle &
             config.
	& \multicolumn{2}{c|}{RHIC} & \multicolumn{2}{c}{LHC(2.76)} & \multicolumn{2}{c}{LHC(5.02)} \\
	& &
	& Sc.1 & Sc.2
	& Sc.1 & Sc.2 
	& Sc.1 & Sc.2 \\
\hline \hline
stat. &
 $f_{0}(980)$ &             --- &\multicolumn{2}{c|}{3.44}   &\multicolumn{4}{c}{6.55} \\
 & $a_{0}(980)$ &             --- &\multicolumn{2}{c|}{10.3}   &\multicolumn{4}{c}{19.6} \\
 & $K_{0}^{\ast}(1430)$ & --- &\multicolumn{2}{c|}{0.738} &\multicolumn{4}{c}{1.18} \\
 & $f_{0}(1770)$ &            --- &\multicolumn{2}{c|}{0.078} &\multicolumn{4}{c}{0.125} \\
 \cline{2-9}
 & $f_{0}(1500)$ &            --- &\multicolumn{2}{c|}{0.239} &\multicolumn{4}{c}{0.403} \\
 \cline{2-9}
 & $\rho(770)$ &               --- &\multicolumn{2}{c|}{94.9}   &\multicolumn{4}{c}{169} \\
 & $\phi(1020)$ &             --- &\multicolumn{2}{c|}{8.49}   &\multicolumn{4}{c}{21.7} \\
 & $\Omega(1672)$ &      --- &\multicolumn{2}{c|}{0.383} &\multicolumn{4}{c}{0.992} \\
\hline \hline
coal. &
 $f_{0}(980)$	&              $\bar{n}n$($P$) & {2.10} & {2.09} & {4.20} & 4.14 & {4.20} & 4.14 \\
 & $a_{0}(980)$	 &             $\bar{n}n$($P$) & {6.31} & {6.28} & {12.62} & 12.42 & {12.62} & 12.42 \\
 & $K_{0}^{\ast}(1430)$	 & $\bar{s}n$($P$) & {0.93} & {0.87} & {1.68} & {1.46} & {1.68} & {1.46} \\
 & $f_{0}(1770)$ &             $\bar{s}s$($P$) & {0.08} & {0.08} & {0.15} & {0.13} & {0.15} & {0.13} \\
 \cline{2-9}
 & $f_{0}(1500)$ &             $\bar{n}n$($P$) & {2.26} & {2.09} & {4.20} & 4.14 & {4.20} & 4.14 \\
 & &                       $\bar{n}^{2}n^{2}$($S$) & {0.08} & {0.09} & {0.14} & 0.19 & {0.14} & 0.19 \\
 & &                                            $gg$($S$) & {0.21} & {0.22} & {0.34} & {0.39} & {0.34} & {0.39} \\ 
 \cline{2-9} 
 & $\rho(770)$ &                $\bar{n}n$($S$) & {80.1} & {80.1} & {159} & 158 & {159} & 158 \\
 & $\phi(1020)$	&               $\bar{s}s$($S$) & 7.34 & 7.31 & 13.1 & 13.1 & 13.1 & 13.1 \\
 & $\Omega(1672)$ &              $sss$($S$) & 0.51 & 0.52 & 0.77 & 0.86 & 0.77 & 0.86 \\
\hline \hline
smat.
 & $f_{0}(500)$ &             --- &\multicolumn{6}{l}{\hspace{0.5em} 15.7 (13.91-16.17)}  \\ 
 & $f_{0}(980)$ &             --- &\multicolumn{6}{l}{\hspace{0.5em} 2.81 (2.03-2.88)}    \\ 
 & $f_{0}(1500)$ &            --- &\multicolumn{6}{l}{\hspace{0.5em} 0.12 (0.043-0.124)}  \\ 
 & $K_{0}^{\ast}(700)$ &             --- &\multicolumn{6}{l}{\hspace{0.5em} 5.373 (4.13-5.52)}   \\ 
 & $K_{0}^{\ast}(1430)$ &          --- &\multicolumn{6}{l}{\hspace{0.5em} 0.570 (0.47-0.58)}   \\ 
 \hline
\end{tabular}
\end{table}

\subsubsection{Yields of hadrons}

We summarize the numerical results for the yields of hadrons per a nucleus collision in the statistical model and the coalescence model in Table~\ref{tbl:numerical_results}.
It shows the results for the new nonet scalar mesons, i.e., $f_{0}(980)$, $a_{0}(980)$, $K_{0}^{\ast}(1430)$, and $f_{0}(1770)$, the glueball candidate, i.e., $f_{0}(1500)$, and the normal hadrons, i.e., $\rho(770)$, $\phi(1020)$, and $\Omega(1672)$.
The yields of hadrons are shown for Sc.1 and Sc.2
in the RHIC and LHC parameters.
It is seen as a general trend that the lighter hadrons are easily produced, while the heavier hadrons are not, as it is intuitively understood by the suppressing mass factor in the particle thermal distribution functions.

As shown in Table~\ref{tbl:numerical_results}, the yields of the new nonet scalar mesons are almost in the range between the order $0.1$ and the order $1$.
They are comparable with the yields of $\rho(770)$, $\phi(1020)$, and $\Omega(1672)$ as reference particles in spite of the different internal angular momenta.
Therefore, it is expected that the observations of the nonet scalar mesons, i.e., $f_{0}(980)$, $a_{0}(980)$, $K_{0}^{\ast}(1430)$, and $f_{0}(1770)$, are feasible in RHIC and LHC experiments.

The yields of the glueball candidate $f_{0}(1500)$ are shown also in Table~\ref{tbl:numerical_results}.
In order to investigate more details, we consider, not only the $gg$ configuration with $S$-wave, but also the other possible internal structures, i.e., quark-antiquark configuration ($\bar{n}n$ with $P$-wave) and tetraquark configuration ($\bar{n}^{2}n^{2}$ with $S$-wave).
As results, it is found that the yields of $f_{0}(1500)$ are much sensitive to the configurations of internal structures.
For example, the yields of the $P$-wave $\bar{n}n$ state is more enhanced than those of the $S$-wave $gg$ state.
This enhancement can be understood from the property that the up and down quark masses in $\bar{n}n$ are lighter than the constituent gluon masses in $gg$.
This enhancement overcomes the $P$-wave suppression factor in the $\bar{n}n$ state.
On the other hand, the yields of the $S$-wave $\bar{n}^{2}n^{2}$ state is more suppressed than those of the $S$-wave $gg$ state.
This suppression can be explained, because the probability for forming a four-particle composite state should be smaller than the probability for the two-particle case when the internal angular momenta are the same.

\subsubsection{Ratios of hadron yields} \label{sec:ratios_hadron_yields}

As mentioned in the Introduction, it is known that the yields of ordinary hadrons in the coalescence model are comparable to those in the statistical model.
Such a comparison can be further clarified by estimating the ratios of hadron yields in the coalescence model and in the statistical model.
This ratio will be denoted by coal./stat. for short.
The results of ratios for Sc.1 in the RHIC parameter are shown in Fig.~\ref{fig:250108_RHIC_Sc1_ratio}.
In the figure, it is found that the yield ratios for normal hadrons, i.e., $\rho(770)$, $\phi(1020)$, and $\Omega(1672)$, are close to one, as it is expected from the previous studies.
The LHC parameters give the similar results.

The yield ratios for the new nonet scalar mesons, i.e., $f_{0}(980)$, $a_{0}(980)$, $K_{0}^{\ast}(1430)$, and $f_{0}(1770)$, and the glueball candidate, i.e., $f_{0}(1500)$, are shown in Fig.~\ref{fig:250108_RHIC_Sc1_ratio}.
These ratios are close to one, although the deviation from one may be relatively more enhanced than the case of normal hadrons.
The yield ratio of $f_{0}(1500)$ estimated for the $S$-wave $gg$ configuration is also close to one.
Therefore, it exhibits the similar behavior as the normal hadrons and the new nonet scalar mesons.
In contrast, the yield ratios of $f_{0}(1500)$ for the $P$-wave $\bar{n}n$ configuration and the $S$-wave $\bar{n}^{2}n^{2}$ configuration provide unusual values, as the deviation from one is much enhanced.
For the gluon candidate $f_{0}(1500)$, thus, we conclude that the $S$-wave $gg$ configuration is more plausible than the other configurations.

Before the conclusion, one may consider that the Casimir factor for $gg$ configuration in $f_{0}(1500)$ can play some role in the coalescence model.
It is known in fact that the Casimir factor in the confinement potential is different between a quark and a gluon.
Thus, the Casimir scaling can change the value of angular momentum frequency for $gg$ in $f_{0}(1500)$.
However, it is shown that introduction of the Casimir factor does not significantly alter the numerical results of the yields of $f_{0}(1500)$.
More details are discussed in Appendix.~\ref{sec:Casimir}.

\section{Conclusions and perspectives}
\label{sec:conclusion_perspectives}

We have developed a novel classification for the scalar nonet, 
$f_{0}(980)$, $a_{0}(980)$, $K_{0}^{\ast}(1430)$, and $f_{0}(1770)$, and have regarded $f_{0}(1500)$ as a glueball.
We have considered their production in HICs, and have estimated their yields by the statistical model and the coalescence model.
The two models produce mutually consistent results.
We have also performed an analysis in terms of the S-matrix formulation, and obtained the consistent results.
It is imperative to validate these findings in future HIC experiments.

Our new nonet scalar mesons and glueballs 
could offer insights into exotic hadrons in the LHC-ALICE experiment,
where it was reported that the $f_{0}(980)$ production is suppressed 
relative to that of $K^{\ast0}(892)$~\cite{ALICE:2023cxn}.
This experimental result suggests that $s\bar{s}$ components in $f_{0}(980)$ should be less than those in the cases of tetraquarks ($n\bar{n}s\bar{s}$) and hadronic molecules ($K\bar{K}$), and consequently, they can be interpreted as $P$-wave $q\bar{q}$ two-quark states.

There remain several topics of interest that were not discussed in the present study.
The scalar meson $f_{0}(1380)$ has been regarded as a kinematically generated state, and hence
we did not consider their production.
If they are indeed formed by the kinematic mechanism, quantifying their production yields will require an intricate scheme different from the simple setting such as the statistical model and the coalescence model.
As discussed in Section II, there can exist another nonet scalar mesons including $f_{0}(1710)$, $a_{0}(1710)$, $K_{0}^{\ast}(1950)$, and $f_{0}(2100)$ and/or $f_{0}(2200)$.
Exploring their structures and the production mechanism in HICs will be of particular importance.
Beyond the standard approaches, 
the study of glueball production in terms of topological solitons
could shed light on the physics of exotic hadrons from a different perspective.
These issues will be addressed in our future studies.

\section*{Acknowledgments}

This work is supported by the World Premier International Research Center Initiative (WPI) in the Ministry of Education, Culture, Sports, Science (MEXT), Japan.
The work of P.~M.~Lo and C.~S. was supported partly by the Polish National Science Centre (NCN) under OPUS Grant No.~2022/45/B/ST2/01527.  The work of S.~H.~Lee was supported by the Korea National Research Foundation under grant No. RS-2023-NR077232.

\appendix

\section{S-matrix calculations}
\label{sec:S-matrix_calculations}

In this appendix we briefly outline the calculation of thermal hadron yields within the S-matrix approach~\cite{dmb,weinhold,nbody,ccm}. A central ingredient of the method is the effective spectral function $B(s)$,

\begin{align}
   B(s) = 2\, \frac{\partial}{\partial \sqrt{s}} \delta(s),
\end{align}
which is derived from the scattering phase shift $\delta(s)$. This effective spectral function quantifies the change in the density of states induced by interactions, whether resonant or non-resonant. When empirical phase shifts from partial-wave analyses are used, the resulting description of the interaction is essentially model independent.

The effective spectral function enters the calculation of thermal yields through
\begin{align}
N_{h}^{\mathrm{S}}
&=
\frac{g_{h} V_{H}}{2\pi^{2}}
\int \frac{\dr \sqrt{s}}{2 \pi} \, B(s)
\int_{0}^{\infty}
\frac{p^{2}\, \dr p}
{e^{(\varepsilon(p,s)-\mu_{B}B-\mu_{S}S)/T_{H}}\pm1},
\end{align}
with $\varepsilon(p,s)=\sqrt{p^2+s}$.
As discussed in the main text, the scattering phase shifts encode the full interaction within a given channel, defined only by its conserved quantum numbers. In this sense, the S-matrix contribution does not correspond uniquely to a specific particle state. This differs from statistical hadronization or coalescence models, where yields are attributed directly to individual hadrons.

To facilitate a direct comparison with those approaches, we evaluate the S-matrix contribution over the invariant-mass interval associated with the state of interest.
For example, we choose the range $2m_{\pi}$ to $0.7$~GeV for the $f_{0}(500)$ and $1.1$--$1.8$~GeV for the $f_{0}(1500)$.
The systematic uncertainty is estimated by varying the chosen mass window.
We adopt a deliberately conservative estimate, leading to substantial uncertainties that can reach the $60$~\% level. These are indicated in parentheses in Table~\ref{tbl:numerical_results}.

\section{Uncertainty in angular momentum frequencies}
\label{sec:discussions}

The coalescence model contains many parameters to be determined appropriately.
In this Appendix, as one of the most unclear parameters, we focus on 
the angular momentum frequencies in the harmonic oscillator potentials:
the $\omega_{s}$ dependence in strange nonet scalar mesons, i.e., $K_{0}^{\ast}(1430)$, and $f_{0}(1770)$, and the Casimir factors in glueballs ($gg$) for $f_{0}(1500)$ (cf.~Sec.~\ref{sec:coalescence} and Sec.~\ref{sec:results}).

\subsection{Angular momentum frequencies in $K_{0}^{\ast}(1430)$ and $f_{0}(1770)$}
\label{sec:ws_dependence}

In Sec.~\ref{sec:coalescence}, we have assumed the enhanced angular momentum frequencies, i.e., $\omega_{s}' \approx 2\omega_{s}$ for $K_{0}^{\ast}(1430)$ and $\omega_{s}'' \approx 3\omega_{s}$ for $f_{0}(1770)$, in the coalescence model.
As already discussed there, these modifications are chosen  such that the production yield of the ground-state hadron matches that obtained in the statistical model.
In this Appendix, we discuss that this modification ansatz for the angular momentum frequencies is indeed needed for realizing the matching condition between the statistical model and the coalescence model.
Besides, we argue that the absence of such modifications may cause some difficulty for interpreting $f_{0}(1500)$ as a glueball.

\begin{figure}[tbp]
\begin{center}
\vspace{0em}
\includegraphics[scale=0.28]{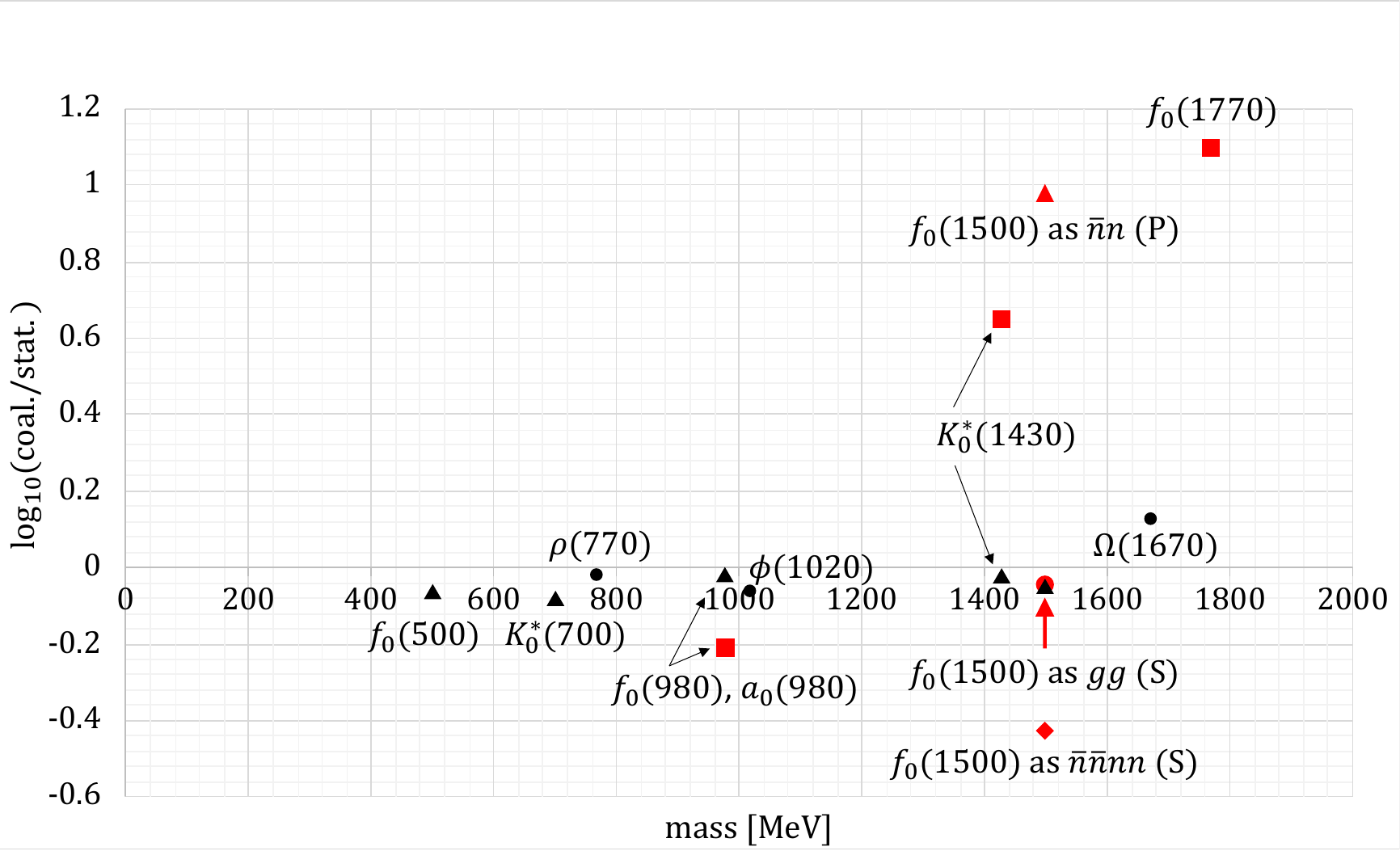}
\caption{The case of non-modification for $\omega_{s}$ in $K_{0}^{\ast}(1430)$ and $f_{0}(1770)$ is presented: $\omega_{s}' = \omega_{s}'' = \omega_{s}$.
The symbols are the same as in Fig.~\ref{fig:250108_RHIC_Sc1_ratio}.
The numerical values for $K_{0}^{\ast}(1430)$ and $f_{0}(1770)$ are shown in Table~\ref{tbl:numerical_results_no_correction}.
This result is compared with Fig.~\ref{fig:250108_RHIC_Sc1_ratio}, where the modified values, $\omega_{s}' \approx 2\omega_{s}$ and $ \omega_{s}'' \approx 3\omega_{s}$, are adopted.}
\label{fig:250207_RHIC_Sc1_ratio_no_corr}
\end{center}
\end{figure}

In order to show the necessity of the modification ansatz,
we suppose tentatively to adopt the non-modified values for $K_{0}^{\ast}(1430)$ and $f_{0}(1770)$: $\omega_{s}' = \omega_{s}'' = \omega_{s}$.
Thus, the angular momentum frequencies are the same as that for $f_{0}(980)$ and $a_{0}(980)$.
In this setting, the results for the yields of $K_{0}^{\ast}(1430)$ and $f_{0}(1770)$ are shown in Table~\ref{tbl:numerical_results_no_correction}, and the ratios of hadron yields are shown in Fig.~\ref{fig:250207_RHIC_Sc1_ratio_no_corr}.
Note that the yields of the other hadrons are not affected by $\omega_{s}'$ and $\omega_{s}''$.
As shown in the figure, the hadron yield ratios for $K_{0}^{\ast}(1430)$ and $f_{0}(1770)$ ($4.4$ and $12.5$ for each) are much different from one, indicating the significant inconsistency between the statistical model and the coalescence model.
From this viewpoint, we confirm that the modification ansatz
provides a sufficient condition for realizing the consistency between the two models.

\begin{table}[tbp]
\caption{The results with non-modification for $\omega_{s}$ for $K_{0}^{\ast}(1430)$ and $f_{0}(1770)$ in $P$-wave, i.e., $\omega_{s}' = \omega_{s}'' = \omega_{s}$, are shown. These results are compared with Table~\ref{tbl:numerical_results}, where the modified values, $\omega_{s}' \approx 2\omega_{s}$ and $ \omega_{s}'' \approx 3\omega_{s}$, are adopted. See Fig.~\ref{fig:250207_RHIC_Sc1_ratio_no_corr} also.}\label{tbl:numerical_results_no_correction}
\begin{tabular}{r|r|r|cc|cccc}
\hline \hline
              &
             particle &
             config.
	& \multicolumn{2}{c|}{RHIC} & \multicolumn{2}{c}{LHC(2.76)} & \multicolumn{2}{c}{LHC(5.02)} \\
	& &
	& Sc.1 & Sc.2
	& Sc.1 & Sc.2 
	& Sc.1 & Sc.2 \\
\hline \hline
coal.
 & $K_{0}^{\ast}(1430)$	 & $\bar{s}n$($P$) & 3.25 & 3.09 & 5.93 & 5.32 & 5.93 & 5.32 \\ %
 & $f_{0}(1770)$ &             $\bar{s}s$($P$) & 0.70 & 0.67 & 1.23 & 1.16 & 1.23 & 1.16 \\ %
\hline
\end{tabular}
\end{table}

One may consider still possibly the following alternative scenario:
the yield ratios of $P$-wave hadrons could exhibit different yields between the statistical model and the coalescence model in essence in contrast to the normal hadron case.
In this case, we cannot completely exclude the possibility that $f_{0}(1500)$ is the $P$-wave $\bar{q}q$ state, because it is allowed that the yields in the coalescence model are different from the statistical model.
This interpretation is, however, not acceptable in the present discussion, because it cannot be consistent with the mass pattern in the new nonet scalar mesons proposed by our study.
In order to be consistent in the whole argument, therefore, we need to regard that the statistical model and the coalescence model are consistent in the $P$-wave hadrons.

\subsection{Casimir factors in glueballs} \label{sec:Casimir}

In the coalescence model in Sec.~\ref{sec:coalescence}, we have adopted the angular momentum frequency $\omega_{g}$ of the harmonic oscillator potential in a glueball ($gg$).
Considering that a gluon belongs to octet representation of the $\mathrm{SU}(3)_{\mathrm{c}}$ color symmetry,
one may wonder whether the Casimir factors play some role for the value of $\omega_{g}$ and hence for the yields of glueballs.
In this Appendix, we discuss the effect by the Casimir factor in the coalescence model in a simple way, and estimate the possible changes of the production yields of $f_{0}(1500)$ as a glueball.

We consider that the linear-type interpotential between a quark ($i$) and an antiquark ($j$) includes the Casimir factor $C_{2} = \vec{\lambda}_{i}\!\cdot\!\vec{\lambda}_{j}$.
Here $\vec{\lambda}_{k}$ are the Gell-Mann matrices operated to the $\mathrm{SU}(3)_{\mathrm{c}}$ fundamental and anti-fundamental representations, i.e., quark and antiquark $k=i,j$, in $q\bar{q}$ case.
This setting is usual in a standard nonrelativistic quark model.
The Casimir factor can be applied also to the $\mathrm{SU}(3)_{\mathrm{c}}$ adjoint representations, i.e., gluons.
The values of $C_{2}$ are $4/3$ for the quark-antiquark ($\bar{q}q$) case and $3$ for the gluon-gluon ($gg$) case.
Thus, we find that the linear-type interpotential strength between two gluons should be enhanced by $9/4$ in comparison to the usual one between a quark and an antiquark.
Supposing that the harmonic oscillator potential is proportional to $C_{2}$, we estimate the change of the $\omega_{g}$ value:
$\omega_{g}$ should be replaced by $(3/2)\omega_{g}$ as the modified angular momentum frequency in confinement potential for gluons.

The modification from $\omega_{g}$ to $(3/2)\omega_{g}$ leads to the change of production yields of $f_{0}(1500)$,
as shown in Table~\ref{tbl:numerical_results_Casimir}.
The yield ratios become smaller ($\approx 0.6$), but they are regarded to be still close to one.
Therefore, we argue that the Casimir factor causes minor changes for the yields of $f_{0}(1500)$ as a glueball.
Interestingly, the yields become closer to the results for tetraquark ($\bar{n}^{2}n^{2}$) in Table~\ref{tbl:numerical_results}.
Although such closeness may be just a coincidence,
it suggests us a possible scenario for the internal structures of $f_{0}(1500)$: the $gg$ configuration and the $\bar{n}^{2}n^{2}$ configuration are different sides of $f_{0}(1500)$.
This hypothetical scenario can be intuitively understood, because a gluon ($g$) and a color-octet quark-antiquark pair ($\bar{n}n$) are interchanged to each other in QCD.
However, it is postponed for future studies to inquire more detail analysis because the confinement potential in the present analysis may be too simple.

\begin{table}[tbp]
\caption{The results with the Casimir factor ($C_{2}$) for glueballs are shown: $\omega_{g}$ are replaced to $(3/2)\omega_{g}$ as the modified confinement potential. The results are compared with the values in Table~\ref{tbl:numerical_results}, where no Casimir factor is adopted.}\label{tbl:numerical_results_Casimir}
\begin{tabular}{r|r|r|cc|cccc}
\hline \hline
              &
             particle &
             config.
	& \multicolumn{2}{c|}{RHIC} & \multicolumn{2}{c}{LHC(2.76)} & \multicolumn{2}{c}{LHC(5.02)} \\
	& &
	& Sc.1 & Sc.2
	& Sc.1 & Sc.2 
	& Sc.1 & Sc.2 \\
\hline \hline
coal. 
 & $f_{0}(1500)$	 & $gg$($S$) & 0.14 & 0.15 & 0.22 & 0.26 & 0.22 & 0.25 \\
\hline
\end{tabular}
\end{table}

\bibliography{reference.bib}

\end{document}